\documentclass[12pt,a4paper]{article}
\usepackage{amsmath}
\usepackage{latexsym}
\makeatletter
\def\rddots{\mathinner{\mkern1mu\raise\p@%
    \vbox{\kern7\p@\hbox{.}}\mkern2mu%
    \raise4\p@\hbox{.}\mkern2mu\raise7\p@\hbox{.}\mkern1mu}}
\makeatother
\newcommand{\ket}[1]{{\vert{#1}\rangle}}

\newcommand{\fukuso}{{\mathbf C}}

\begin{document}

\title{\sl Rotating Wave Approximation of the Law's Effective 
Hamiltonian on the Dynamical Casimir Effect}
\author{
  Kazuyuki FUJII
  \thanks{E-mail address : fujii@yokohama-cu.ac.jp }\quad and\ \ 
  Tatsuo SUZUKI
  \thanks{E-mail address : suzukita@sic.shibaura-it.ac.jp }\\
  ${}^{*}$International College of Arts and Sciences\\
  Yokohama City University\\
  Yokohama, 236--0027\\
  Japan\\
  ${}^{\dagger}$Department of Mathematical Sciences\\
  College of Systems Engineering and Science\\
  Shibaura Institute of Technology\\
  Saitama, 337--8570\\
  Japan\\
  }
\date{}
\maketitle
\begin{abstract}
  In this paper we treat the Law's effective Hamiltonian of 
the Dynamical Casimir Effect in a cavity and construct 
an analytic approximate solution of the time--dependent 
Schr{\"o}dinger equation under the general setting 
through a kind of rotating wave approximation (RWA). 
To the best of our knowledge this is the finest analytic 
approximate solution.
\end{abstract}

\vspace{5mm}\noindent

\section{Introduction}
In this paper we revisit the so-called dynamical Casimir effect (DCE). 
It means the photon generation from vacuum due to the motion 
(change) of neutral boundaries, which corresponds to a kind of 
quantum fluctuation of the electro-magnetic field. 

This phenomenon is a typical example of interactions between 
the microscopic and the macroscopic levels and is very fascinating 
from the point of view of not only (pure) Physics but also 
Mathematical Physics. 
As a general introduction to this topic, see for example \cite{Dodonov}.

In the paper we treat the effective Hamiltonian by Law \cite{Law} 
(which is time--dependent) and propose a kind of rotating wave 
approximation to the model and construct an analytic 
approximate solution under the general setting in terms of 
some infinite dimensional representation of the Lie algebra $su(1,1)$. 
This is the best solution known so far.

\section{Model and RWA}
First of all let us make a brief review of Law \cite{Law} within 
our necessity. See also \cite{Dodonovs} and \cite{FS-3}. 
In the following we set $\hbar=1$ for simplicity. 

As an effective Hamiltonian of the dynamical Casimir effect 
in a cavity (Cavity DCE for simplicity) 
we adopt the simplest one which is the special 
case of \cite{Law} (namely, $\epsilon(x,t)=\epsilon(t)$). 
However, this is fruitful enough as shown in the following. We set
\begin{equation}
\label{eq:Law}
H=\omega(t)a^{\dagger}a+i\chi(t)\left\{(a^{\dagger})^{2}-a^{2}\right\}
\end{equation}
where $\omega(t)$ is a periodic function depending on the cavity form 
and $\chi(t)$ is given by
\[
\chi(t)
=\frac{1}{4\omega(t)}\frac{d\omega(t)}{dt}
=\frac{1}{4}\frac{d}{dt}\log|\omega(t)|.
\]
Here $a$ and $a^{\dagger}$ are the cavity photon annihilation and 
creation operators respectively. 
Therefore, the physics that we are treating is two-photon generation 
processes from the vacuum state.

We would like to solve the Schr{\"o}dinger equation
\begin{equation}
\label{eq:Schrodinger}
i\frac{d}{dt}\ket{\Psi(t)}=H\ket{\Psi(t)}=H(t)\ket{\Psi(t)}
\end{equation}
explicitly under the general setting.

If $H$ is time--independent, then the general (formal) solution 
is given by
\[
\ket{\Psi(t)}=e^{-itH}\ket{\Psi(0)}.
\]
Of course, to calculate $e^{-itH}$ exactly is another problem 
(which is in general very hard).  
However, in our case $H$ is time--dependent, so 
solving (\ref{eq:Schrodinger}) becomes increasingly difficult. 
We must make some approximation to the Hamiltonian.

We impose further restrictions on the model. Since 
$\omega(t)$ is periodic we take $\omega(t)$ as in \cite{Dodonovs} 
\[
\omega(t)=\omega_{0}(1+\epsilon\sin(\eta t))
\]
where $\omega_{0}$, $\epsilon$ and $\eta$ are real constants. 
We assume that $\omega_{0}>0$, $0<\epsilon\ll 1$ and 
$\eta$ is large enough. 
Under the restrictions we may set
\[
\omega(t)\approx \omega_{0}\quad \mbox{and}\quad
\chi(t)
=\frac{\epsilon\eta\cos(\eta t)}{4(1+\epsilon\sin(\eta t))}
\approx \frac{\epsilon\eta}{4}\cos(\eta t).
\]
Then the Hamiltonian (\ref{eq:Law}) is approximated by
\begin{equation}
\label{eq:Law-2}
H=\omega_{0}a^{\dagger}a+
i\frac{\epsilon\eta}{4}\cos(\eta t)\left\{(a^{\dagger})^{2}-a^{2}\right\}.
\end{equation}

We apply a kind of rotating wave approximation (RWA) to this model. 
Since
\[
\cos(\eta t)
=\frac{e^{i\eta t}+e^{-i\eta t}}{2}
=\frac{1}{2}e^{i\eta t}\left(1+e^{-2i\eta t}\right)
=\frac{1}{2}e^{-i\eta t}\left(1+e^{2i\eta t}\right)
\]
we neglect the term $e^{-2i\eta t}$ or $e^{2i\eta t}$ like
\[
\cos(\eta t)\left\{(a^{\dagger})^{2}-a^{2}\right\}
=\cos(\eta t)(a^{\dagger})^{2}-\cos(\eta t)a^{2}
\approx \frac{1}{2}\left\{e^{-i\eta t}(a^{\dagger})^{2}-e^{i\eta t}a^{2}\right\}.
\]
Note that the right hand side is anti--hermitian. 
See for example \cite{WS}, \cite{JC} for more details on RWA.

As a result our Hamiltonian (\ref{eq:Law-2}) changes to
\begin{equation}
\label{eq:Law-3}
\widetilde{H}
=\omega_{0}a^{\dagger}a+
i\frac{\epsilon\eta}{8}\left\{e^{-i\eta t}(a^{\dagger})^{2}-e^{i\eta t}a^{2}\right\}
=\omega_{0}N+
i\frac{\epsilon\eta}{8}\left\{e^{-i\eta t}(a^{\dagger})^{2}-e^{i\eta t}a^{2}\right\},
\end{equation}
which is hermitian. Our aim is to solve the modified Schr{\"o}dinger equation
\begin{equation}
\label{eq:Schrodinger-modified}
i\frac{d}{dt}\ket{\Psi(t)}=\widetilde{H}\ket{\Psi(t)}=\widetilde{H}(t)\ket{\Psi(t)}
\end{equation}
explicitly under the general setting.

\section{Method of Solution}
In order to solve the equation (\ref{eq:Schrodinger-modified}) we use 
a well--known Lie algebraic method, see for example \cite{KF}, 
\cite{Kazuyuki}. If we set
\begin{equation}
\label{eq:su(1,1)-generators}
K_{+}=\frac{1}{2}(a^{\dagger})^{2},\quad
K_{-}=\frac{1}{2}a^{2},\quad
K_{3}=\frac{1}{2}\left(N+\frac{1}{2}\right)
\end{equation}
it is not difficult to see both $K_{+}^{\dagger}=K_{-},\ K_{3}^{\dagger}=K_{3}$ 
and the $su(1,1)$ relations
\begin{equation}
\label{eq:su(1,1)-relations}
[K_{3},K_{+}]=K_{+},\quad [K_{3},K_{-}]=-K_{-},\quad [K_{+},K_{-}]=-2K_{3}
\end{equation}
by use of the relations (Heisenberg relations)
\[
[N,a^{\dagger}]=a^{\dagger},\quad [N,a]=-a,\quad 
[a,a^{\dagger}]=1\ \ (\Longleftrightarrow [a^{\dagger},a]=-1).
\]
In terms of $\{K_{+},K_{-},K_{3}\}$ the Hamiltonian (\ref{eq:Law-3}) 
can be written as
\begin{equation}
\label{eq:Law-4}
\widetilde{H}=-\frac{\omega_{0}}{2}+
2\omega_{0}K_{3}+\frac{i\epsilon\eta}{4}\left(e^{-i\eta t}K_{+}-e^{i\eta t}K_{-}\right).
\end{equation}

Next, we review the basic $su(1,1)$ relations. If we set $\{k_{+},k_{-},k_{3}\}$ as
\begin{equation}
\label{eq:basic su(1,1) generators}
 k_{+} = \left(
        \begin{array}{cc}
               0 & 1 \\
               0 & 0 
        \end{array}
       \right),
 \quad 
 k_{-} = \left(
        \begin{array}{cc}
               0 & 0 \\
              -1 & 0 
        \end{array}
       \right),
 \quad 
 k_{3} = \frac12 
       \left(
        \begin{array}{cc}
               1 &  0 \\
               0 & -1 
        \end{array}
       \right)
\end{equation}
it is easy to check the basic $su(1,1)$ relations
\begin{equation}
\label{eq:basic su(1,1)-relations}
[k_{3},k_{+}]=k_{+},\quad [k_{3},k_{-}]=-k_{-},\quad [k_{+},k_{-}]=-2k_{3}.
\end{equation}
Note that $k_{+}^{\dagger}=-k_{-}\ (\Leftrightarrow 
K_{+}^{\dagger}=K_{-}) $. 
That is, $\{k_{+},k_{-},k_{3}\}$ are generators of the Lie algebra 
$su(1,1)$ of the non-compact group $SU(1,1)$.

Since $SU(1,1)$ is contained in the special linear group $SL(2;\fukuso)$, 
we {\bf assume} that there exists an infinite dimensional unitary 
representation 
$\rho : SL(2;\fukuso)\ \longrightarrow\ U({\cal F})$ 
(group homomorphism\footnote{In general, to construct a (Lie) group 
homomorphism is very hard. See some textbook of Lie groups.}) satisfying
\begin{equation}
\label{eq:correspondence}
d\rho(k_{+})=K_{+},\quad d\rho(k_{-})=K_{-},\quad d\rho(k_{3})=K_{3}
\end{equation}
where $d\rho$ is its differential representation and ${\cal F}$ is the Fock 
space generated by the generators $\{a^{\dagger},a,N\equiv a^{\dagger}a\}$ 
stated above (see \cite{Kazuyuki} for more details).

Then, since $d\rho$ is linear the main part of (\ref{eq:Law-4}) becomes
\begin{eqnarray*}
2\omega_{0}K_{3}+\frac{i\epsilon\eta}{4}\left(e^{-i\eta t}K_{+}-e^{i\eta t}K_{-}\right)
&=&
2\omega_{0}d\rho(k_{3})+
\frac{i\epsilon\eta}{4}\left(e^{-i\eta t}d\rho(k_{+})-e^{i\eta t}d\rho(k_{-})\right) \\
&=&
d\rho
\left(
2\omega_{0}k_{3}+\frac{i\epsilon\eta}{4}\left(e^{-i\eta t}k_{+}-e^{i\eta t}k_{-}\right)
\right)
\end{eqnarray*}
and we set
\begin{equation}
\widetilde{h}\equiv 
2\omega_{0}k_{3}+\frac{i\epsilon\eta}{4}\left(e^{-i\eta t}k_{+}-e^{i\eta t}k_{-}\right)
=
\left(
\begin{array}{cc}
\omega_{0} & \frac{i\epsilon\eta}{4}e^{-i\eta t}  \\
\frac{i\epsilon\eta}{4}e^{i\eta t} & -\omega_{0}
\end{array}
\right).
\end{equation}
Note that $\widetilde{h}$ is not hermitian. 

Here, we would like to solve the (small) Schr{\"o}dinger--like equation
\begin{equation}
\label{eq:small Schrodinger}
i\frac{d}{dt}\ket{\psi(t)}=\widetilde{h}\ket{\psi(t)}=\widetilde{h}(t)\ket{\psi(t)}.
\end{equation}
For the purpose we decompose $\widetilde{h}=\widetilde{h}(t)$ into
\[
\left(
\begin{array}{cc}
\omega_{0} & \frac{i\epsilon\eta}{4}e^{-i\eta t}  \\
\frac{i\epsilon\eta}{4}e^{i\eta t} & -\omega_{0}
\end{array}
\right)
=
\left(
\begin{array}{cc}
e^{-i\frac{\eta}{2}t} &   \\
  & e^{i\frac{\eta}{2}t} 
\end{array}
\right)
\left(
\begin{array}{cc}
\omega_{0} & \frac{i\epsilon\eta}{4}   \\
\frac{i\epsilon\eta}{4} & -\omega_{0}
\end{array}
\right)
\left(
\begin{array}{cc}
e^{i\frac{\eta}{2}t} &     \\
  & e^{-i\frac{\eta}{2}t} 
\end{array}
\right)
\]
and set
\[
\ket{\phi(t)}=
\left(
\begin{array}{cc}
e^{i\frac{\eta}{2}t} &     \\
  & e^{-i\frac{\eta}{2}t} 
\end{array}
\right)
\ket{\psi(t)}
\Longleftrightarrow 
\ket{\psi(t)}=
\left(
\begin{array}{cc}
e^{-i\frac{\eta}{2}t} &   \\
  & e^{i\frac{\eta}{2}t} 
\end{array}
\right)
\ket{\phi(t)}.
\]
Then, it is easy to see
\begin{equation}
\label{eq:modified small Schrodinger}
i\frac{d}{dt}\ket{\phi(t)}
=
\left(
\begin{array}{cc}
\omega_{0}-\frac{\eta}{2} & \frac{i\epsilon\eta}{4}                  \\
\frac{i\epsilon\eta}{4} & -\left(\omega_{0}-\frac{\eta}{2}\right)
\end{array}
\right)
\ket{\phi(t)}
\equiv A\ket{\phi(t)}.
\end{equation}
Since $A$ is time--independent the general solution is 
given by
\[
\ket{\phi(t)}=e^{-itA}\ket{\phi(0)},
\]
and to calculate the term $e^{-itA}$ is relatively easy.

For simplicity we set
\[
A=
\left(
\begin{array}{cc}
\omega_{0}-\frac{\eta}{2} & \frac{i\epsilon\eta}{4}                   \\
\frac{i\epsilon\eta}{4} & -\left(\omega_{0}-\frac{\eta}{2}\right)
\end{array}
\right)
\equiv 
\left(
\begin{array}{cc}
c & id   \\
id & -c 
\end{array}
\right),
\quad 
c=\omega_{0}-\frac{\eta}{2},\ d=\frac{\epsilon\eta}{4}.
\]
Noting
\[
A^{2}=(c^{2}-d^{2})1_{2}, \quad
c^{2}-d^{2}=
\left(\omega_{0}-\frac{\eta}{2}\right)^{2}-\left(\frac{\epsilon\eta}{4}\right)^{2}
\]
(we assume that $c>d$) we have
\begin{eqnarray}
\label{eq:exponential calculation}
e^{-itA}
&=&
\sum_{n=0}^{\infty}\frac{(-it)^{2n}}{(2n)!}A^{2n}+
\sum_{n=0}^{\infty}\frac{(-it)^{2n+1}}{(2n+1)!}A^{2n+1} \nonumber \\
&\equiv&
\left(
\begin{array}{cc}
\bar{\alpha} & \beta  \\
\beta & \alpha
\end{array}
\right)
\end{eqnarray}
where
\[
\alpha=\alpha(t)=\cos\left(t\sqrt{c^{2}-d^{2}}\right)+
i\frac{\sin\left(t\sqrt{c^{2}-d^{2}}\right)}{\sqrt{c^{2}-d^{2}}}c, \quad
\beta=\beta(t)=\frac{\sin\left(t\sqrt{c^{2}-d^{2}}\right)}{\sqrt{c^{2}-d^{2}}}d
\]
and
\[
\frac{\beta}{\alpha}
=
\frac
{
\frac{\sin\left(t\sqrt{c^{2}-d^{2}}\right)}{\sqrt{c^{2}-d^{2}}}d
}{
\cos\left(t\sqrt{c^{2}-d^{2}}\right)+i\frac{\sin\left(t\sqrt{c^{2}-d^{2}}\right)}{\sqrt{c^{2}-d^{2}}}c
}.
\]

\vspace{5mm}
Going back to $\ket{\psi(t)}$ from $\ket{\phi(t)}$ we obtain
\begin{equation}
\ket{\psi(t)}
=
\left(
\begin{array}{cc}
e^{-i\frac{\eta}{2}t} &   \\
  & e^{i\frac{\eta}{2}t} 
\end{array}
\right)
\left(
\begin{array}{cc}
\bar{\alpha} & \beta  \\
\beta & \alpha
\end{array}
\right)\ket{\psi(0)}
=
\left(
\begin{array}{cc}
\bar{\alpha}e^{-i\frac{\eta}{2}t} & \beta e^{-i\frac{\eta}{2}t}  \\
\beta e^{i\frac{\eta}{2}t} & \alpha e^{i\frac{\eta}{2}t} 
\end{array}
\right)\ket{\psi(0)}
\end{equation}
with (\ref{eq:exponential calculation}). Here, note that
\[
\left|
\begin{array}{cc}
\bar{\alpha}e^{-i\frac{\eta}{2}t} & \beta e^{-i\frac{\eta}{2}t}  \\
\beta e^{i\frac{\eta}{2}t} & \alpha e^{i\frac{\eta}{2}t} 
\end{array}
\right|
=
|\alpha|^{2}-\beta^{2}=1
\Longrightarrow 
\left(
\begin{array}{cc}
\bar{\alpha}e^{-i\frac{\eta}{2}t} & \beta e^{-i\frac{\eta}{2}t}     \\
\beta e^{i\frac{\eta}{2}t} & \alpha e^{i\frac{\eta}{2}t} 
\end{array}
\right) \in SL(2;{\bf C}).
\]

\vspace{3mm}
Next, let us recall the (well--known) Gauss decomposition of 
elements in $SL(2;{\bf C})$ :
\[
  \left(
   \begin{array}{cc}
     a & b \\
     c & d \\
   \end{array}
  \right)
=
  \left(
   \begin{array}{cc}
     1 & \frac{b}{d}  \\
     0 & 1
   \end{array}
  \right)
  \left(
   \begin{array}{cc}
     \frac{1}{d} & 0  \\
     0 & d
   \end{array}
  \right)
  \left(
   \begin{array}{cc}
     1 & 0             \\
     \frac{c}{d} & 1
   \end{array}
  \right)\quad (d\ne 0,\ ad-bc=1)
\]
because
\[
\frac{1}{d}+\frac{bc}{d}=\frac{1+bc}{d}=\frac{ad}{d}=a.
\]

This formula gives
\begin{eqnarray*}
\left(
\begin{array}{cc}
\bar{\alpha}e^{-i\frac{\eta}{2}t} & \beta e^{-i\frac{\eta}{2}t}     \\
\beta e^{i\frac{\eta}{2}t} & \alpha e^{i\frac{\eta}{2}t} 
\end{array}
\right)
&=&
\left(
\begin{array}{cc}
1 & \frac{\beta}{\alpha}e^{-i\eta t}  \\
0 & 1 
\end{array}
\right)
\left(
\begin{array}{cc}
\frac{1}{\alpha}e^{-i\frac{\eta}{2}t} & 0  \\
0 & \alpha e^{i\frac{\eta}{2}t} 
\end{array}
\right)
\left(
\begin{array}{cc}
1 & 0                                  \\
\frac{\beta}{\alpha} & 1 
\end{array}
\right)  \\
&=&
\exp
\left(
\begin{array}{cc}
0 & \frac{\beta}{\alpha}e^{-i\eta t}  \\
0 & 0 
\end{array}
\right)\times \\
&{}&
\exp
\left(
\begin{array}{cc}
-\log\alpha-i\frac{\eta}{2}t & 0  \\
0 & \log\alpha+i\frac{\eta}{2}t 
\end{array}
\right)\times \\
&{}&
\exp
\left(
\begin{array}{cc}
0 & 0                           \\
\frac{\beta}{\alpha} & 0 
\end{array}
\right) \\
&=&
\exp\left(\frac{\beta}{\alpha}e^{-i\eta t}k_{+}\right)
\exp\left(-2\left(\log\alpha+i\frac{\eta}{2}t\right)k_{3}\right)
\exp\left(-\frac{\beta}{\alpha}k_{-}\right)
\end{eqnarray*}
by (\ref{eq:basic su(1,1) generators}).

\vspace{3mm}
Here, we apply the (group) homomorphism $\rho$ to this case. 
Since $\rho$ is a group homomorphism (namely, 
$\rho(ABC)=\rho(A)\rho(B)\rho(C)$) we have

\begin{eqnarray*}
&&\rho\left(
\left(
\begin{array}{cc}
\bar{\alpha}e^{-i\frac{\eta}{2}t} & \beta e^{-i\frac{\eta}{2}t}     \\
\beta e^{i\frac{\eta}{2}t} & \alpha e^{i\frac{\eta}{2}t} 
\end{array}
\right)
\right) \\
&=&
\rho\left(\exp\left(\frac{\beta}{\alpha}e^{-i\eta t}k_{+}\right)\right)
\rho\left(\exp\left(-2\left(\log\alpha+i\frac{\eta}{2}t\right)k_{3}\right)\right)
\rho\left(\exp\left(-\frac{\beta}{\alpha}k_{-}\right)\right)  \\
&=&
\exp\left(\frac{\beta}{\alpha}e^{-i\eta t}d\rho(k_{+})\right)
\exp\left(-2\left(\log\alpha+i\frac{\eta}{2}t\right)d\rho(k_{3})\right)
\exp\left(-\frac{\beta}{\alpha}d\rho(k_{-})\right)\ \ (\Downarrow \mbox{by definition})  \\
&=&
\exp\left(\frac{\beta}{\alpha}e^{-i\eta t}K_{+}\right)
\exp\left(-2\left(\log\alpha+i\frac{\eta}{2}t\right)K_{3}\right)
\exp\left(-\frac{\beta}{\alpha}K_{-}\right)\ \ (\Downarrow \mbox{by}\ (\ref{eq:correspondence})).
\end{eqnarray*}

This is just the disentangling formula that we are looking for. 
Note that our derivation is heuristic (not logical) because we have 
assumed the existence of the Lie group homomorphism $\rho$, 
which is not established.

We can {\bf conjecture} the general solution of the 
equation (\ref{eq:Schrodinger-modified}) with (\ref{eq:Law-4}) 
as
\begin{equation}
\label{eq:conjecture}
\ket{\Psi(t)}=e^{i\frac{\omega_{0}}{2}t}
\exp\left(\frac{\beta}{\alpha}e^{-i\eta t}K_{+}\right)
\exp\left(-2\left(\log\alpha+i\frac{\eta}{2}t\right)K_{3}\right)
\exp\left(-\frac{\beta}{\alpha}K_{-}\right)
\ket{\Psi(0)}.
\end{equation}

\noindent
Let us prove this conjecture. For the purpose we set
\begin{equation}
U(t)=e^{i\frac{\omega_{0}}{2}t}e^{f(t)K_{+}}e^{g(t)K_{3}}e^{h(t)K_{-}}, 
\quad U(0)=1\ (\mbox{identity})
\end{equation}
with unknown functions $f(t), g(t), h(t)\ (f(0)=g(0)=h(0)=0)$. 
By use of the method developed in \cite{Kazuyuki} or \cite{FS-3} 
we obtain
\begin{equation}
i\frac{d}{dt}U(t)=
\left\{
-\frac{\omega_{0}}{2}+
i(\dot{f}-\dot{g}f+\dot{h}e^{-g}f^{2})K_{+}+
i(\dot{g}-2\dot{h}e^{-g}f)K_{3}+
i\dot{h}e^{-g}K_{-}
\right\}U(t)
\end{equation}
where we have used the notations like $\frac{df}{dt}=\dot{f}$, etc 
for simplicity. 

If we set
\begin{eqnarray}
f(t)
&=&\frac{\beta}{\alpha}e^{-i\eta t}=
\frac
{
\frac{\sin\left(t\sqrt{c^{2}-d^{2}}\right)}{\sqrt{c^{2}-d^{2}}}d
}{
\cos\left(t\sqrt{c^{2}-d^{2}}\right)+i\frac{\sin\left(t\sqrt{c^{2}-d^{2}}\right)}{\sqrt{c^{2}-d^{2}}}c
}e^{-i\eta t},  \nonumber \\
g(t)
&=&-2\left(\log\alpha+i\frac{\eta}{2}t\right)
=
-2\left\{
\log\left(\cos\left(t\sqrt{c^{2}-d^{2}}\right)+
i\frac{\sin\left(t\sqrt{c^{2}-d^{2}}\right)}{\sqrt{c^{2}-d^{2}}}c\right)
+i\frac{\eta}{2}t
\right\},  \nonumber \\
h(t)
&=&-\frac{\beta}{\alpha}
=
-\frac
{
\frac{\sin\left(t\sqrt{c^{2}-d^{2}}\right)}{\sqrt{c^{2}-d^{2}}}d
}{
\cos\left(t\sqrt{c^{2}-d^{2}}\right)+i\frac{\sin\left(t\sqrt{c^{2}-d^{2}}\right)}{\sqrt{c^{2}-d^{2}}}c
}
\end{eqnarray}
with $c=\omega_{0}-\frac{\eta}{2}$ and $d=\frac{\epsilon\eta}{4}$, \ 
then a long but straightforward calculation shows
\begin{eqnarray*}
&&-\frac{\omega_{0}}{2}+
i(\dot{f}-\dot{g}f+\dot{h}e^{-g}f^{2})K_{+}+
i(\dot{g}-2\dot{h}e^{-g}f)K_{3}+
i\dot{h}e^{-g}K_{-}  \\
&=&
-\frac{\omega_{0}}{2}+\frac{i\epsilon\eta}{4}e^{-i\eta t}K_{+}
+2\omega_{0}K_{3}-\frac{i\epsilon\eta}{4}e^{i\eta t}K_{-} \\
&=&
-\frac{\omega_{0}}{2}+
2\omega_{0}K_{3}+\frac{i\epsilon\eta}{4}\left(e^{-i\eta t}K_{+}-e^{i\eta t}K_{-}\right) \\
&=&\widetilde{H}.
\end{eqnarray*}
As a result the general solution is given by
\begin{equation}
\label{eq:finish}
U(t)\ket{\Psi(0)}=\ket{\Psi(t)}.
\end{equation}
This concludes the proof.

\vspace{3mm}\noindent
For example, when $\ket{\Psi(0)}=\ket{0}$ the vacuum state 
($a\ket{0}=0$), some calculation by use of (\ref{eq:su(1,1)-generators}) 
gives
\begin{eqnarray}
\label{eq:example}
\ket{\Psi(t)}
&=&e^{i\frac{\omega_{0}}{2}t}
\exp\left(\frac{\beta}{\alpha}e^{-i\eta t}K_{+}\right)
\exp\left(-2\left(\log\alpha+i\frac{\eta}{2}t\right)K_{3}\right)
\exp\left(-\frac{\beta}{\alpha}K_{-}\right)
\ket{0}  \nonumber \\
&=&\frac{e^{i\frac{c}{2}t}}{\sqrt{\alpha}}
\sum_{n=0}^{\infty}\sqrt{{}_{2n}C_{n}}\left(\frac{\beta}{2\alpha}e^{-i\eta t}\right)^{n}
\ket{2n}
\end{eqnarray}
with $\alpha=\alpha(t),\beta=\beta(t)$ and $c=\omega_{0}-\frac{\eta}{2}$ 
in (\ref{eq:exponential calculation}).

\section{Reconfirmation}
In this section let us reconfirm our method. 
Our real target is the following ``Hamiltonian"
\[
h=
\left(
\begin{array}{cc}
\omega_{0} & \frac{i\epsilon\eta}{2}\cos(\eta t)   \\
\frac{i\epsilon\eta}{2}\cos(\eta t) & -\omega_{0}
\end{array}
\right)
\]
($h$ is not hermitian) and to solve the (small) Schr{\"o}dinger--like 
equation
\[
i\frac{d}{dt}\ket{\psi(t)}=h\ket{\psi(t)}=h(t)\ket{\psi(t)}.
\]
However, it is very difficult to solve at the present time. 

Since
\[
\left(
\begin{array}{cc}
e^{i\frac{\eta}{2}t} &    \\
 & e^{-i\frac{\eta}{2}t}
\end{array}
\right)
h
\left(
\begin{array}{cc}
e^{-i\frac{\eta}{2}t} &   \\
 & e^{i\frac{\eta}{2}t}
\end{array}
\right)
=
\left(
\begin{array}{cc}
\omega_{0} & \frac{i\epsilon\eta}{4}   \\
\frac{i\epsilon\eta}{4} & -\omega_{0}
\end{array}
\right)
+
\left(
\begin{array}{cc}
  & \frac{i\epsilon\eta}{4}e^{2i\eta t}  \\
\frac{i\epsilon\eta}{4}e^{-2i\eta t} & 
\end{array}
\right)
\]
we can neglect the last term by the rotating wave approximation 
when $\eta$ is large enough. 
Therefore, we changed $h$ into $\widetilde{h}$
\[
\widetilde{h}=
\left(
\begin{array}{cc}
e^{-i\frac{\eta}{2}t} &   \\
 & e^{i\frac{\eta}{2}t}
\end{array}
\right)
\left(
\begin{array}{cc}
\omega_{0} & \frac{i\epsilon\eta}{4}   \\
\frac{i\epsilon\eta}{4} & -\omega_{0}
\end{array}
\right)
\left(
\begin{array}{cc}
e^{i\frac{\eta}{2}t} &    \\
 & e^{-i\frac{\eta}{2}t}
\end{array}
\right)
=
\left(
\begin{array}{cc}
\omega_{0} & \frac{i\epsilon\eta}{4}e^{-i\eta t}  \\
\frac{i\epsilon\eta}{4}e^{i\eta t} & -\omega_{0}
\end{array}
\right)
\]
($\widetilde{h}$ is still not hermitian) and solved 
the equation
\[
i\frac{d}{dt}\ket{\psi(t)}=\widetilde{h}\ket{\psi(t)}=\widetilde{h}(t)\ket{\psi(t)}
\]
explicitly.

\section{Concluding Remarks}
In this paper we treated the Law's effective Hamiltonian 
of the Dynamical Casimir Effect in a cavity, and considered  
a kind of rotating wave approximation for the model, and 
constructed an analytic approximate solution under any 
initial condition. 
Our construction is based on some infinite dimensional 
representation of the Lie algebra $su(1,1)$. 
We believe that the method is simple, powerful and beautiful. 
As for related topics see \cite{EFS}, \cite{FS} and 
\cite{FS-1}, \cite{FS-2} (the last two are highly recommended). 

Our approximate Hamiltonian is still time--dependent. 
In general, to solve the Schr{\"o}dinger equation with a
time--dependent Hamiltonian explicitly is very hard. 
As stated in the abstract this is the best analytic approximate 
solution as far as we know. 

What we did in the paper is of course nothing but an intermediate 
stage because we must detect photon generated by the vacuum state. 
How do we detect ?  How do we construct a unified model 
containing the detection ? 
Such a model has been presented by \cite{Dodonovs} and 
\cite{FS-3}. In a forthcoming paper (papers) we will apply 
the result in this paper to the model.

\vspace{10mm}\noindent 
{\bf Acknowledgments}\\
We would like to thank Ryu Sasaki for useful suggestions and comments.



\begin{thebibliography}{99}
%
\bibitem{Dodonov}V. V. Dodonov : 
\newblock Current status of the Dynamical Casimir Effect, 
\newblock Physica Scripta, {\bf 82} (2010), 038105.
\newblock arXiv : 1004.3301 [quant-ph].
%
\bibitem{Law}C. K. Law : 
\newblock Effective Hamiltonian for the radiation in a cavity 
with a moving mirror and a time--varying dielectric medium, 
\newblock Phys. Rev. A {\bf 49} (1994), 309.
%
\bibitem{Dodonovs}A. V. Dodonov and V. V. Dodonov : 
\newblock Approximate analytical results on the cavity Casimir 
effect in the presence of a two--level atom, 
\newblock Phys. Rev. A {\bf 85} (2012), 063804.
\newblock arXiv : 1112.0523 [quant-ph].
%
\bibitem{FS-3}K. Fujii and T. Suzuki :
\newblock An Approximate Solution of the Dynamical Casimir 
Effect in a Cavity with a Two-Level Atom, 
\newblock Int. J. Geom. Methods Mod. Phys, {\bf 10} (2013), 1350035, 
\newblock arXiv : 1209.5133 [quant-ph].
%
\bibitem{WS}W. P. Schleich : 
\newblock Quantum Optics in Phase Space,
\newblock WILEY--VCH, Berlin, 2001.
%
\bibitem{JC}E. T. Jaynes and F. W. Cummings : 
\newblock Comparison of Quantum and Semiclassical Radiation
Theories with Applications to the Beam Maser, 
\newblock Proc. IEEE, {\bf 51} (1963), 89.
%
\bibitem{KF}K. Fujii : 
\newblock Introduction to Coherent States and Quantum Information Theory, 
\newblock quant-ph/0112090.
%
\bibitem{Kazuyuki}K. Fujii : 
\newblock Quantum Damped Harmonic Oscillator, 
\newblock Chapter 7 of ``Advances in Quantum Mechanics",
Paul Bracken (Ed.), ISBN 980-953-307-945-0, InTech, 
\newblock arXiv:1209.1437 [quant-ph].
%
\bibitem{EFS}R. Endo, K. Fujii and T. Suzuki :
\newblock General Solution of the Quantum Damped Harmonic Oscillator, 
\newblock Int. J. Geom. Meth. Mod. Phys, {\bf 5} (2008), 653, 
\newblock arXiv : 0710.2724 [quant-ph].
%
\bibitem{FS}K. Fujii and T. Suzuki :
\newblock General Solution of the Quantum Damped Harmonic Oscillator II :
Some Examples, 
\newblock Int. J. Geom. Meth. Mod. Phys, {\bf 6} (2009), 225, 
\newblock arXiv : 0806.2169 [quant-ph].
%
\bibitem{FS-1}K. Fujii and T. Suzuki :
\newblock An Approximate Solution of the Jaynes--Cummings 
Model with Dissipation, 
\newblock Int. J. Geom. Methods Mod. Phys, {\bf 8} (2011), 1799, 
\newblock arXiv : 1103.0329 [math-ph].
%
\bibitem{FS-2}K. Fujii and T. Suzuki :
\newblock An Approximate Solution of the Jaynes--Cummings 
Model with Dissipation II : Another Approach, 
\newblock Int. J. Geom. Methods Mod. Phys, {\bf 9} (2012), 1250036, 
\newblock arXiv : 1108.2322 [math-ph].
\end{thebibliography}
\end{document}